\documentclass[a4paper]{jpconf}
\usepackage{latexsym}
\usepackage{graphicx}
\begin{document}
\title{Magnetic phase diagram of Sr$_{1-x}$Ca$_x$Co$_2$P$_2$}

\author{J Sugiyama$^1$, H Nozaki$^1$, I Umegaki$^1$, M Harada$^1$, Y~Higuchi$^1$, 
E~J~Ansaldo$^2$, J H Brewer$^{2,3}$, M Imai$^4$, C Michioka$^4$, K Yoshimura$^4$, and M M\aa nsson$^{5,6}$}

\address{$^1$ Toyota Central Research \& Development Laboratories, Inc., 41-1 Yokomichi, Nagakute, Aichi 480-1192, Japan}
\address{$^2$ TRIUMF, 4004 Wesbrook Mall, Vancouver, BC, V6T 2A3 Canada}
\address{$^3$ Department of Physics \& Astronomy, University of British Columbia, Vancouver, BC, V6T 1Z1 Canada}
\address{$^4$ Department of Chemistry, Graduate School of Science, Kyoto Univ., Kyoto, 606-8502 Japan}
\address{$^5$ Laboratory for Quantum Magnetism, $\acute{\rm E}$cole Polytechnique F$\acute{\rm e}$d$\acute{\rm e}$rale de Lausanne, 
CH-1015, Switzerland}
\address{$^6$ Laboratory for Neutron Scattering, Paul Scherrer Institut, CH-5232 Villigen PSI, Switzerland}

\ead{e0589@mosk.tytlabs.co.jp}

\begin{abstract}
In order to study the phase diagram from a microscopic viewpoint, 
we have measured wTF- and ZF-$\mu^+$SR spectra for the Sr$_{1-x}$Ca$_x$Co$_2$P$_2$ powder samples with 
$x=0$, 0.2, 0.4, 0.5, 0.6, 0.8, and 1. 
Due to a characteristic time window and spatial resolution of $\mu^+$SR, 
the obtained phase diagram was found to be rather different from that determined by magnetization measurements. 
That is, as $x$ increases from 0, 
a Pauli-paramagnetic phase is observed even at the lowest $T$ measured (1.8~K) until $x=0.4$,  
then, a spin-glass like phase appears at $0.5\leq x\leq0.6$, 
and then, a phase with wide field distribution probably due to incommensurate AF order is detected for $x=0.8$, 
and finally, a commensurate $A$-type AF ordered phase (for $x=1$) is stabilized below $T_{\rm N}\sim80~$K. 
Such change is most likely reasonable and connected to the shrink of the $c$-axis length with $x$, 
which naturally enhances the magnetic interaction between the two adjacent Co planes.  

\end{abstract}

\section{Introduction}

The appearance of unconventional superconductivity in iron pnictides  
with the ThCr$_2$Si$_2$-type (122) structure, i.e. 
Ba$_{1-x}$K$_x$Fe$_2$As$_2$ reconfirmed 
the competition between antiferromagnetism and superconductivity \cite{rotter_08a,rotter_08b}.  
Similar to the K-doping effect,  
the magnetic state of CaFe$_2$As$_2$ varies with pressure from antiferromagnetic (AF) to superconducting, 
and finally to nonmagnetic \cite{torikachvili_08}, 
while its structure changes from an ``uncollapsed tetragonal" (uc$T$) phase, 
an orthorhombic phase, 
and a ``collapsed tetragonal" (c$T$) phase \cite{kreyssig_08}. 
Here, in the collapsed phase,  
the ratio of stacking to in-plane lattice parameters ($c/a$) is significantly smaller than 
that expected from simple atomic size considerations \cite{reehuis_90}.

In contrast to CaFe$_2$As$_2$ and BaFe$_2$As$_2$, 
the related compounds, $AM_2$P$_2$ with $A=~$Ca, Sr, and Ba, and $M=~$Fe, Co, and Ni 
do not show unconventional superconductivity, 
but exhibit an interesting magnetic transition with the structural change. 
For the present target system, Sr$_{1-x}$Ca$_x$Co$_2$P$_2$, 
the crystal structure changes 
from uc$T$ for $x=0$ to c$T$ for $x=1$ \cite{jia_08a}. 
According to the electronic and magnetic phase diagram 
determined from the macroscopic measurements \cite{jia_08a},  
the system evolves from a nonmagnetic metallic ground state to an AF metallic ground state through a
crossover composition regime at $x\sim0.5$. 
Then, in the $x$ range between 0.8 and 0.9, the system manifests a FM-like ground state 
within which the magnetic ordering temperature is highest at $x\sim0.9$. 
For the sample with $x\geq0.9$, an AF ground state reappears \cite{reehuis_98}. 
A similar behavior was also reported for Ca(Fe$_{1-x}$Co$_x$)$_2$P$_2$ \cite{jia_10}, Ca(Ni$_{1-x}$Co$_x$)$_2$P$_2$ \cite{jia_10}, 
and SrCo$_2$(Ge$_{1-x}$P$_x$)$_2$ \cite{jia_11}.

Note that $\chi$ measurements usually give us a significant insight into 
the ground state of magnetically ordered solids. 
However,    
such measurements are sometimes not suitable, 
particularly for the materials exhibiting order with a broad field distribution, 
{\it i.e.} when short-range order, random, nearly-random order, or incommensurate order appears in a material, 
due to the absence of periodic structure 
and/or the presence of rapid fluctuations.  
We have therefore performed a $\mu^+$SR experiment on Sr$_{1-x}$Ca$_x$Co$_2$P$_2$ 
in order to investigate the variation of the magnetic nature with the Ca content ($x$). 

\section{Experimental}

\begin{figure}[hbt]
\begin{minipage}{18pc}
\includegraphics[width=\columnwidth]{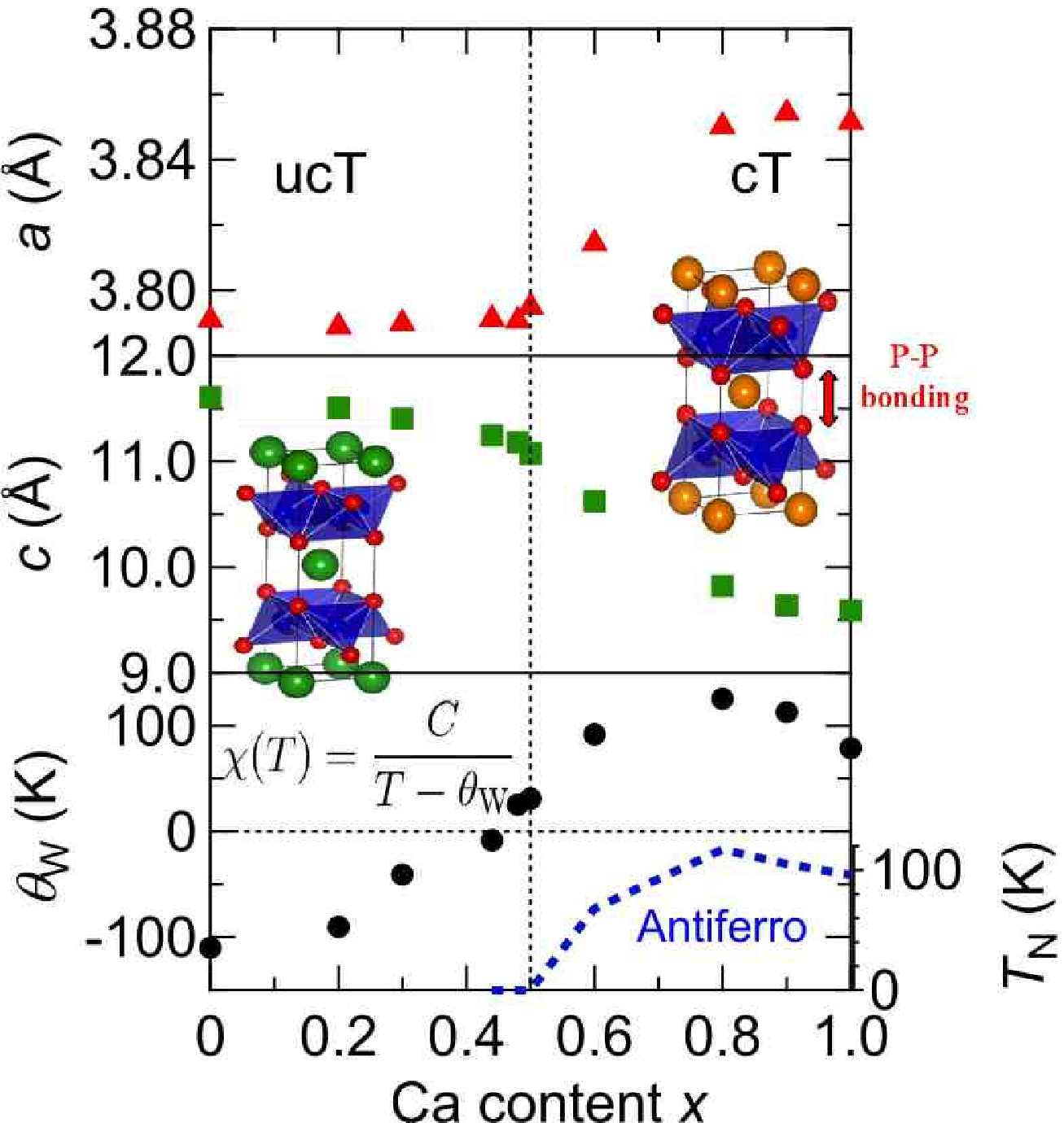}
\caption{\label{fig:xrdchi}
The change in the lattice constants and magnetic properties of Sr$_{1-x}$Ca$_x$Co$_2$P$_2$ with $x$ 
determined by XRD analyses and $\chi$ measurements \cite{imai}. 
}
\end{minipage}\hspace{2pc}%
\begin{minipage}{18pc}
\includegraphics[width=\columnwidth]{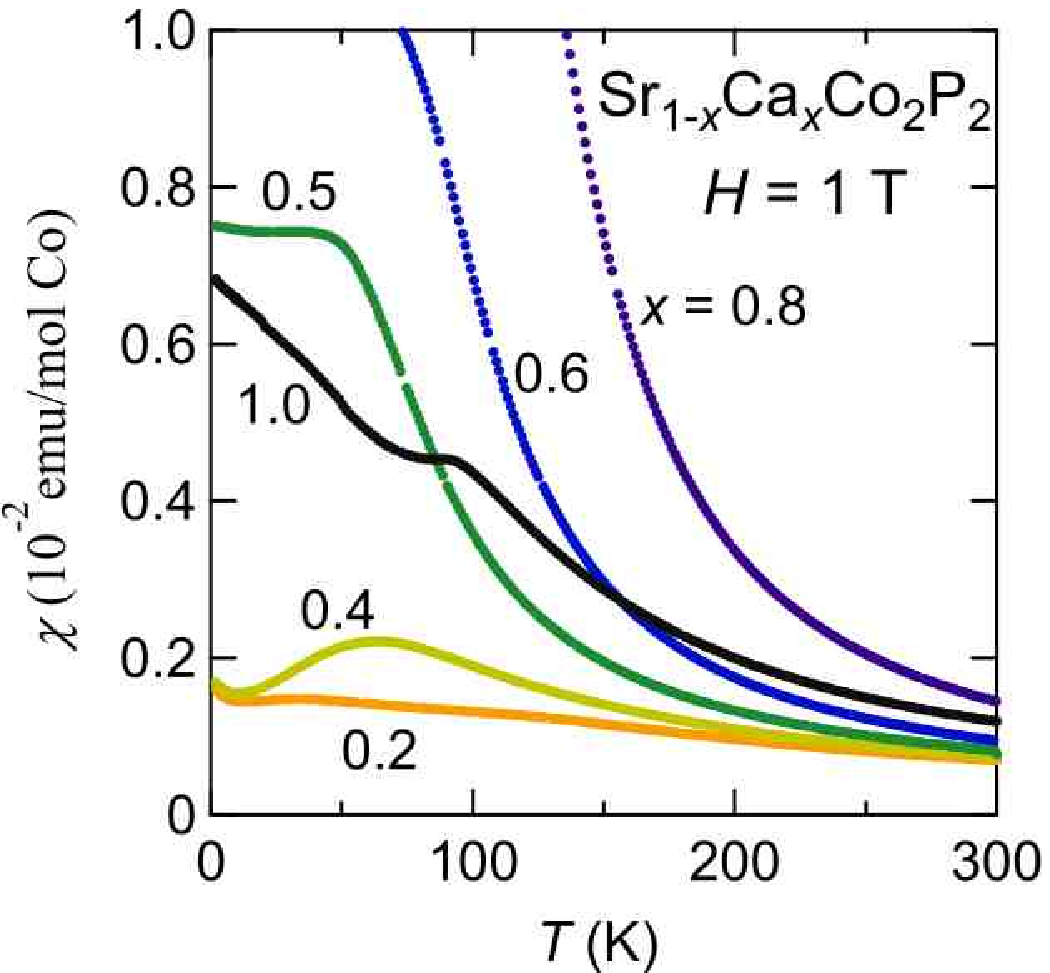}
\caption{\label{fig:ZFana}
$T$ dependence of $\chi$ for Sr$_{1-x}$Ca$_x$Co$_2$P$_2$ with $x\geq0.2$ \cite{imai}.
}
\end{minipage} 
\end{figure}

Polycrystalline samples of Sr$_{1-x}$Ca$_x$Co$_2$P$_2$ with $x=0$, 0.2, 0.4, 0.5, 0.6, 0.8, and 1 were prepared 
from elemental P, Sr, Ca, and Co using two step reaction.
For the first step, SrP, CaP, and Co$_2$P were synthesized by a solid state reaction 
between Sr (Ca, Co) and P in an evacuated quartz tube at 800$^{\circ}$C (700$^{\circ}$C for Co$_2$P). 
Then, for the second step, Sr$_{1-x}$Ca$_x$Co$_2$P$_2$ were synthesized by a solid state reaction 
between SrP, CaP and Co$_2$P at 1000$^{\circ}$C for 20 hours in an Ar atmosphere. 
After grinding, the obtained powder was fired two times at 1000$^{\circ}$C for 40 hours in an Ar atmosphere \cite{imai_12,imai}. 

According to powder x-ray diffraction (XRD) analyses,  
all the samples were single phase of tetragonal symmetry  
with space group $I4/mmm$. 
Also, as $x$ increases from 0, 
the $c$-axis length monotonically decreases with $x$ until $\sim0.5$, 
then rapidly decreases until 0.9,  
and finally, levels off to a constant value above $x=0.9$. 
This behavior is consistent with the literature \cite{jia_08a}.  
Susceptibility ($\chi$) measurements suggested the presence of magnetic transition above $\sim60~$K 
for the samples with $x\geq0.6$. 

The $\mu^+$SR spectra were measured
at surface muon beam lines using the {\bf LAMPF} spectrometer of M20/TRIUMF in Canada.  
An approximately 500~mg powder sample was placed in 
an envelope with $1\times1$~cm$^2$ area, 
made with Al-coated Mylar tape
with 0.05~mm thickness 
in order to minimize the signal from the envelope.
Then, the envelope was attached to a low-back ground sample holder 
in a liquid-He flow-type cryostat for measurements in the $T$ range between 1.8 and 150~K. 

\section{Results and discussion}

\begin{figure}[b]
\begin{minipage}{18pc}
\includegraphics[width=\columnwidth]{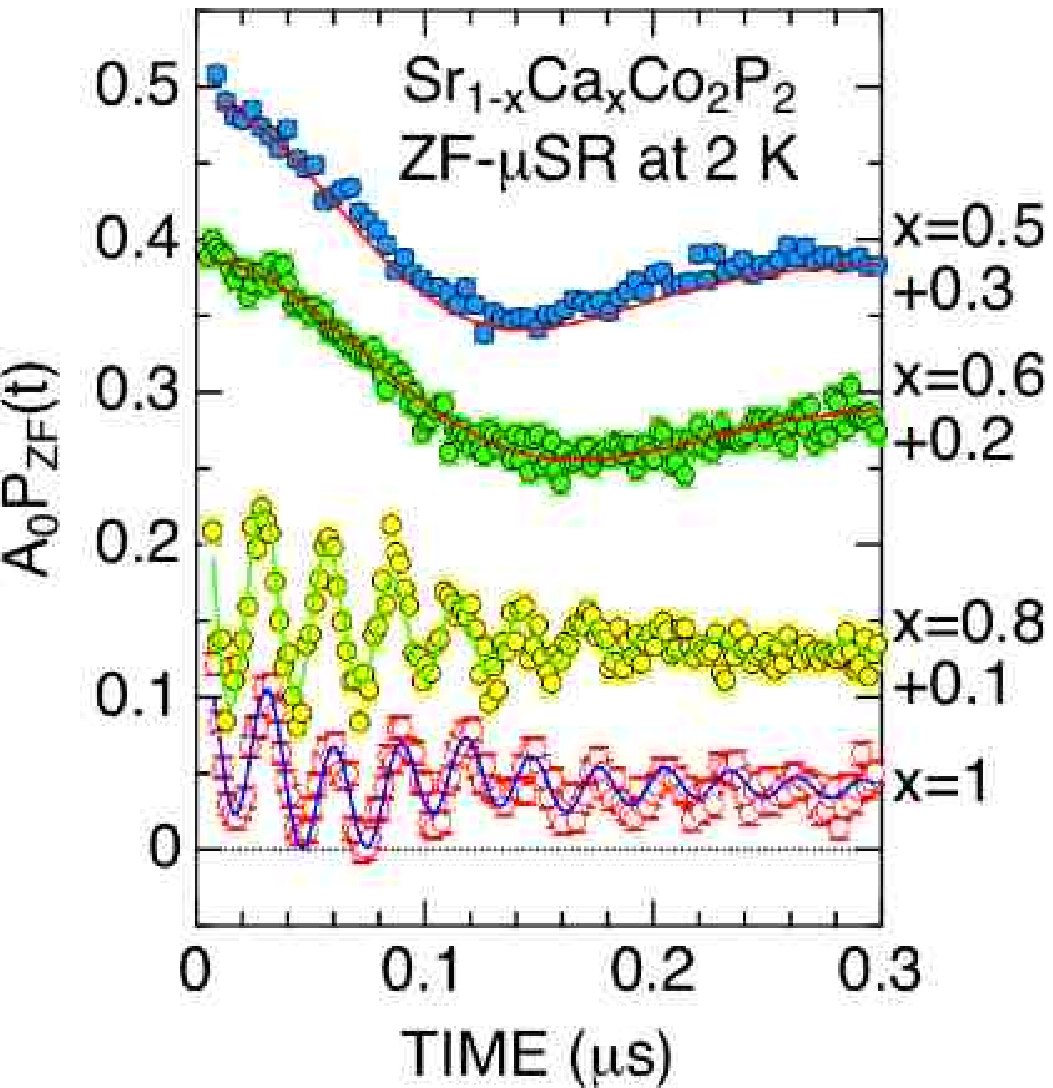}
\caption{\label{fig:ZFspectra}
The ZF-$\mu^+$SR spectra at 2~K for the Sr$_{1-x}$Ca$_x$Co$_2$P$_2$ sample with $x=0$, 0.2, 0.4, and 0.5.
}
\end{minipage}\hspace{2pc}%
\begin{minipage}{18pc}
\includegraphics[width=\columnwidth]{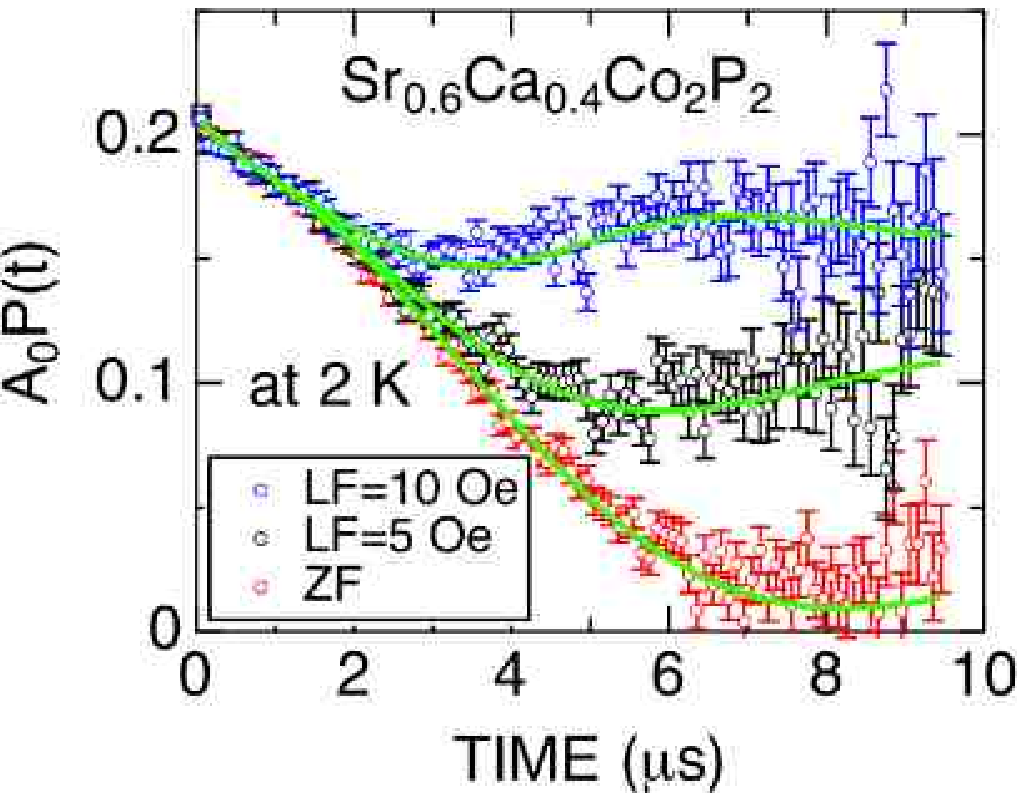}
\caption{\label{fig:ZFLF}
The ZF- and LF-$\mu^+$SR spectra at 2~K for the Sr$_{0.6}$Ca$_{0.4}$Co$_2$P$_2$ sample. 
The applied LF was 5 and 10 Oe. 
Solid lines represent the fit result using a static Gaussian Kubo-Toyabe function. 
By fitting both ZF- and LF-spectra using a common field distribution width ($\Delta$), 
we can obtain reliable $\Delta$.
}
\end{minipage} 
\end{figure}

Figure~\ref{fig:ZFspectra} shows the ZF-$\mu^+$SR spectrum 
for Sr$_{1-x}$Ca$_x$Co$_2$P$_2$ with $x=0.5$, 0.6, 0.8, and 1 
at the lowest $T$ measured. 
The ZF-spectrum for CaCo$_2$P$_2$ exhibits a clear oscillation 
due to the formation of static AF order below $T_{\rm N}\sim80~$K, 
as reported by neutron diffraction measurements \cite{reehuis_98}. 
However, according to the detailed analysis, it was found that the ZF-spectrum consists of two muon-spin precession signals 
with different frequencies [$f_{\rm AF1}=34.6(1)~$MHz and $f_{\rm AF2}=9.5(3)~$MHz at 2~K]. 
As $T$ increases from 2~K, the two signals disappeared at $T_{\rm N}$. 
This suggests that there are two magnetically different muon sites in the CaCo$_2$P$_2$ lattice. 
 
Although a clear oscillation is also observed in the ZF-spectrum for Sr$_{0.2}$Ca$_{0.8}$Co$_2$P$_2$, 
the initial phase ($\phi$) was found to delay by $12-60^{\circ}$, 
when we attempted to fit the oscillatory signal by an exponential relaxing cosine function: 
$A_0 P_{\rm ZF}(t)=A_{\rm AF}\cos(2\pi f_{\rm AF}t+\phi)$. 
This implies the wide distribution of an internal field at the muon site(s). 
The most probable reason is that the magnetic order is incommensurate (IC) to the lattice \cite{kalvius}.  
This is because, due to the mismatch of the period between the muon site(s) and magnetic order, 
IC order always provides an oscillatory signal with wide field distribution in the ZF-$\mu^+$SR spectrum. 
But, the other explanation is also available, as in the case for Ag$_2$NiO$_2$ \cite{jun_06} and LiCrO$_2$ \cite{jun_09}. 
Therefore, neutron diffraction measurements is required for further clarifying the magnetic nature of the $x=0.8$ sample. 

For Sr$_{0.5}$Ca$_{0.5}$Co$_2$P$_2$, the ZF-spectrum shows a static Kubo-Toyabe-type relaxation. 
Since the field distribution width ($\Delta$) was estimated as $12\times10^6~$s$^{-1}(\sim141~$Oe), 
such KT behavior is naturally caused by disordered Co moments. 
Furthermore, weal transverse field $\mu^+$SR measurements 
revealed the appearance of a large internal magnetic field below $\sim50~$K.  
Therefore, the ground state of Sr$_{0.5}$Ca$_{0.5}$Co$_2$P$_2$ is most likely a spin-glass like state, 
although it is more preferable to measure AC-$\chi$ to confirm the spin-glass state.
The situation for Sr$_{0.4}$Ca$_{0.6}$Co$_2$P$_2$ was the same to that for Sr$_{0.5}$Ca$_{0.5}$Co$_2$P$_2$. 

On the other hand, for the samples with $x\leq0.4$, 
the ZF-spectrum exhibits an almost static KT behavior even at 2~K (Fig.~\ref{fig:ZFLF}), but $\Delta=0.22\times10^6~$s$^{-1}(\sim2.6~$Oe). 
This is a typical value for nuclear magnetic fields at the muon site(s), 
and for this case, the origin is mainly due to the nuclear magnetic moment of $^{59}$Co. 
Hence, the ground state for Sr$_{1-x}$Ca$_x$Co$_2$P$_2$ with $x\leq0.4$ is assigned to be a Pauli-paramagnet. 

\begin{figure}[h]
\begin{center}
\includegraphics[width=85 mm]{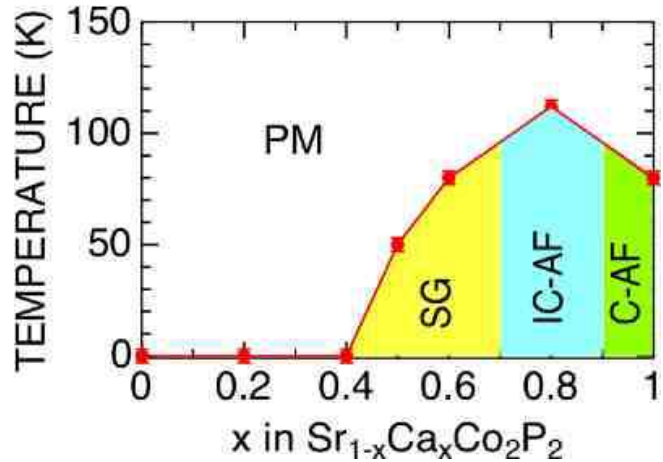}
\caption{\label{fig:phase}
The tentative phase diagram for Sr$_{1-x}$Ca$_x$Co$_2$P$_2$ determined by $\mu^+$SR, 
where PM is a paramagnetic phase, 
SG is a spin-glass like phase, 
IC-AF is a phase with wide field distribution probably due to incommensurate AF order, and 
C-AF is a commensurate AF phase.  
}
\end{center}
\end{figure}

Based on these results, the magnetic phase diagram for Sr$_{1-x}$Ca$_x$Co$_2$P$_2$ was determined (Fig.~\ref{fig:phase}). 
That is, thanks to a unique power of $\mu^+$SR, 
the phase with $x\sim0.8$ was found to be not an FM phase \cite{jia_08a} but an incommensurate (IC) AFM phase. 
Then, next to the IC-AFM phase, a spin-glass (SG) like phase appears in the $x$ range between $\sim0.7$ and $\sim0.4$. 
The samples with $x\leq0.4$ was found to be nonmagnetic down to 2~K.
However, since the phase boundaries between C and IC, between IC and SG, and SG and PM are still not clarified. 
we need additional measurements on several Sr$_{1-x}$Ca$_x$Co$_2$P$_2$ samples. 

\section{Acknowledgments}
We thank the staff of TRIUMF    
for help with the $\mu^+$SR experiments.
This work was supported by MEXT KAKENHI Grant No. 23108003 and 
JSPS KAKENHI Grant No. 26286084.


\section*{References}
\begin{thebibliography}{9}

\bibitem{rotter_08a} Rotter M, Tegel M, Johrendt D, Schellenberg I, Hermes W, and P$\ddot{\rm o}$ttgen R 2008
{\it Phys. Rev.} B {\bf 78} 020503
%
\bibitem{rotter_08b} Rotter M, Tegel M, and Johrendt D 2008 
{\it Phys. Rev. Lett.} {\bf 101} 107006 
%
\bibitem{torikachvili_08} Torikachvili M S, Bud'ko S L, Ni N, and Canfield P C 2008 
{\it Phys. Rev. Lett.} {\bf 101} 057006
%
\bibitem{kreyssig_08} Kreyssig A, Green M A, Lee Y, Samolyuk G D, Zajde P, Lynn J W, Bud'ko S L, Torikachvili M S, Ni N, Nandi S, Le$\tilde{\rm a}$o J B, Poulton S J, Argyriou D N, Harmon B N, McQueeney R J, Canfield P C, and Goldman A I 2008  
{\it Phys. Rev.} B {\bf 78} 184517
%
\bibitem{reehuis_90} Reehuis M and Jeitschko W 1990 
{\it J. Phys. Chem. Solids} {\bf 51} 961
%
\bibitem{jia_08a} Jia S, Williams A J, Stephens P W, and Cava R J 209 
{\it Phys. Rev.} B {\bf 80} 165107
%
\bibitem{reehuis_98} Reehuis M, Jeitschko W, Kotzyba G, Zimmer B, and Hu X 1998
{\it J. Alloys Compounds} {\bf 266} 54 
%
\bibitem{jia_10} Jia S, Chi S, Lynn J W, and Cava R J 2010
{\it Phys. Rev.} B {\bf 81} 214446 
%
\bibitem{jia_11} Jia S, Jiramongkolchai P, Suchomel M R, Toby B H, Checkelsky J G, Ong N P, and Cava R J 2011
{\it Nature Phys.} {\bf 7} 207
%
\bibitem{imai_12} Imai M 2012 {\it Syntheses and itinerant-electron magnetic properties
of the layered cobalt phosphide $A$Co$_2$P$_2$ ($A =$~Ca, Sr, Ba, La)}
(Kyoto: Kyoto University)
%
\bibitem{imai} Imai M, Michioka C, Ohta H, Matsuo A, Kindo K, Ueda H, and Yoshimura K unpublished.
%
\bibitem{kalvius} Kalvius G M, Noakes D R, and Hartmann O 2001 
{\it Handbook on the Physics and Chemistry of Rare Earths} (Amsterdam: North-Holland) vol 32 chapter 206 pp 55--451
%
\bibitem{jun_06} Sugiyama J, Ikedo Y, Mukai K, Brewer J H, Ansaldo E J, Chow K H, Yoshida H, and Hiroi Z 2006  
{\it Phys. Rev.} B {\bf 73}, 224437
%
\bibitem{jun_09} Sugiyama J, M\aa nsson M, Ikedo Y, Goko T, Mukai K, Andreica D, Amato A, Ariyoshi K, and Ohzuku T 2009 
{\it Phys. Rev.} B {\bf 79}, 184411
%


\end{thebibliography}
\end{document}